\documentclass[12pt,a4paper]{article}
\usepackage[]{amsmath,amssymb}
\usepackage{graphics,epsfig}
\begin{document}
\date{}
\title{Wedge reflection positivity}
\author{H. Casini\footnote{e-mail: casini@cab.cnea.gov.ar}\\
{\sl Centro At\'omico Bariloche and Instituto Balseiro}\\ 
{\sl 8400-S.C. de Bariloche, R\'{\i}o Negro, Argentina}}
\maketitle
\begin{abstract}
We show there is a positivity property for Wightman functions which is analogous to the reflection positivity for the euclidean ones. The role of euclidean time reflections is played here by the wedge reflections, which change the sign of the time and one of the spatial coordinates. 
\end{abstract}
\section{Introduction}
The positivity of the Hilbert space scalar product gives place to an infinite series of inequalities involving the correlation functions of any number of variables in a quantum field theory (QFT). In the real-time formulation we have, for any finite sequence of test functions
 $f_0\in \mathbb{C}$, $f_1(x_1)$, ..., $f_k(x_1,...,x_k)$, the inequality 
\begin{equation}
\sum_{i,j=0}^k \int \,dx_1 ...dx_i \, dy_1... dy_j \, (f_i(x_1,...,x_i))^* {\cal W}_{i+j}(x_i,...,x_1,y_1,...,y_j) f_j(y_1,...,y_j)\ge 0\,,\label{wig} 
\end{equation}
where ${\cal W}_{n}(x_1,...,x_n)$ are the Wightman distributions (for a real scalar field) \cite{w1}. This property allows for the reconstruction of the Hilbert space from the correlation functions and plays a central role in the Wightman axiomatic framework \cite{sw}. In (\ref{wig}) the integrations are over all the $d$-dimensional Minkowski space, $d\ge 2$. These inequalities always involve the singularities of the Wightman functions at coinciding points, i.e. when $x_i=y_j$. Therefore, the content of (\ref{wig}) for a finite number of functions at a finite number of points is far from being transparent, and its consequences are sometimes more easily seen in momentum space.     

A different situation holds in the euclidean framework \cite{os,os1}.  The relation corresponding to (\ref{wig}) in this case is called reflection positivity, and writes    
\begin{equation}
\sum_{i,j=0}^k \int \,dx_1...dx_i \, dy_1... dy_j \, (f_i(x_1,...,x_i))^* {\cal E}_{i+j}
(\hat{x}_i,...,\hat{x}_1,y_1,...,y_j) f_j(y_1,...,y_j)\ge 0\,. \label{rp} 
\end{equation}
Here ${\cal E}_{n}
(x_1,...,x_n)$ are the Schwinger functions (euclidean correlators), the integration is over the euclidean space and $\hat{x}=(-x^0,x^1,...,x^{d-1})$ is the euclidean time-reflected point corresponding to $x=(x^0,x^1,...,x^{d-1})$. The test functions $f_j(x_1,...,x_j)$ have support only for the time ordered points on the positive-time half-space $0<x_1^0<...<x_j^0$.  Thus, the inequalities (\ref{rp}), in contrast to the ones in real time (\ref{wig}), only involve non coinciding points for the correlators.  The reflection positivity property provides the connection between euclidean statistical interpretation of the Schwinger functions and the quantum interpretation in terms of a relativistic QFT.

The purpose of this paper is to show the Wightman functions satisfy a positivity relation resembling reflection positivity. This seems to have escaped previous attention. In order to introduce these inequalities let us define the wedge in Minkowski space as the open set $\mathbb{W}=\{x\in \mathbb{R}^d;\, x^1> \vert x^0 \vert\}$. This is bounded by the two null planes intersecting on the $(d-2)$-dimensional spatial plane $\{x\in \mathbb{R}^d;\, x=(0,0,x^2,...,x^{d-1})  \}$. 
We define an order relation on the points of $\mathbb{R}^d$ as
 $x\triangleleft y$ iff $y-x \in \mathbb{W}$. In particular, if $x\triangleleft y$ holds, $x$ and $y$ are space-like separated. The 
 wedge reflection positivity  (WRP) relations for a hermitian scalar field read
\begin{equation}
\sum_{i,j=0}^k \int \,dx_1 ...dx_i \, dy_1... dy_j \, (f_i(x_1,...,x_i))^* {\cal W}_{i+j}
(\bar{x}_i,...,\bar{x}_1,y_1,...,y_j) f_j(y_1,...,y_j)\ge 0\,. \label{wedgereflection} 
\end{equation}
Here the wedge reflection is $\bar{x}=(-x^0,-x^1,x^2,...,x^{d-1})$, and the inequalities hold for any finite sequence of test functions $f_0$, $f_1(x_1)$, ..., $f_k(x_1,...,x_k)$, 
where $f_j(x_1,...,x_j)$ can be non zero only if the points $x_1$,...,$x_j$ are wedge ordered, $0\triangleleft x_1\triangleleft...\triangleleft x_j$.

The WRP is naturally understood as a consequence of the Tomita-Takesaki theory for the algebra of operators on the wedge, revealing the TCP theorem has an associated positivity property.  However, the inequalities (\ref{wedgereflection}) are valid in greater generality. We prove them from the positivity, covariance and spectral properties of Wightman functions, without using the TCP theorem, or, equivalently, weak local commutativity \cite{ot}.   

The WRP does not involve the correlators at coinciding points. Then, specific inequalities for a finite number of Wightman functions evaluated at definite points can be derived. Given any collection of wedge ordered sets of points, $A_i=\{x_1^{(i)},...,x_{n_i}^{(i)}\}$, $0\triangleleft x_1^{(i)}\triangleleft...\triangleleft x_{n_i}^{(i)}$, for $i=1,...,m$, we have from (\ref{wedgereflection}), taking the limit of localized test functions,
\begin{equation}
\textrm{det}\left(\{{\cal W}(\bar{A}_{i}A_j)\}_{i,j=1...m}\right)\ge 0\,,\label{cuatro}
\end{equation}
where we write $\bar{A}_i=\{\bar{x}_{n_i}^{(i)},...,\bar{x}_{1}^{(i)}\}$.
 
Recently, we have shown that the exponentials $e^{(n-1)I_n(\bar{A}_i , A_j)}$ of the Renyi mutual information $I_n(\bar{A}_i,A_j)$ of integer index $n$, between disjoint regions bounded by the sets of points $\bar{A}_i$ and $A_j$ in two-dimensional QFT, obeys the inequalities (\ref{cuatro}) \cite{ca}. The Renyi entropies measure essentially the entanglement of the vacuum state. The inequalities indicate they are given by the vacuum expectation values of some local operators. This is also indicated by the path integral representations of the Renyi entropies in the euclidean framework \cite{cc}. An interesting application of the WRP inequalities would be to show the validity of this mapping between vacuum entanglement and field operators with full rigor. We postpone the study of a reconstruction theorem in this sense, based on the WRP inequalities, to a future work.

\section{Proof of wedge reflection positivity}

In order to make the exposition more clear, let us consider the case of a hermitian scalar field first, and then show the necessary changes for the case of fields with spin. In order to prove WRP we use analyticity of the Wightman functions. The proof is very similar in form to the one of reflection positivity for the Schwinger functions. 

Let us first introduce some notation, which closely follows the one in \cite{os}. The open future light cone is $V^+=\{x:\, x.x> 0,\, x^0>0\}$, where $x.x=(x^{0})^2-\sum_{i=1}^{d-1} (x^i)^2$ is the Minkowski scalar product, and we call the closed cone $V^+_c$. We write the Minkowski metric $G=\textrm{diag}(1,-1,...,-1)$. Let ${\cal S}(\mathbb{R}^{d n})$ be the Schwartz space of infinite differentiable complex test functions of fast decrease on $\mathbb{R}^{d n}$, with the usual
 topology. Using the notation for the partial derivatives 
$D^\alpha=\partial^{\vert\alpha\vert}/((\partial x_1^0)^{\alpha_1} ...(\partial x_n^{d-1})^{\alpha_{(d-1)n}})$, where $\alpha=(\alpha_1,...,\alpha_{(d-1)n})$ we also define the following closed subspaces of 
${\cal S}(\mathbb{R}^{d n})$  
\begin{eqnarray}
{\cal S}_+(\mathbb{R}^{dn})&=&\{   f\in  {\cal S}(\mathbb{R}^{d n}); D^\alpha f (x_1,...,x_n)=0\,\,\, \forall \alpha  \,\,\, \textrm{unless} \,\,\,\,  0\triangleleft x_1\triangleleft ... \triangleleft x_n \} \,,\\
 {\cal S}_\triangleleft(\mathbb{R}^{dn})&=&\{   f\in  {\cal S}(\mathbb{R}^{d n}); D^\alpha f (x_1,...,x_n)=0\,\,\, \forall \alpha  \,\,\, \textrm{unless} \,\,\,\,  x_1\triangleleft...\triangleleft x_n         \}\,.
\end{eqnarray}
For each test function space ${\cal S}$ we call ${\cal S}^\prime$ the corresponding dual space of distributions. 

Let us start the proof by recalling some well-known facts. Because of translation invariance we have 
\begin{equation}
{\cal W}_n(x_1,x_2,...,x_n)=W_{n-1}(\xi_1,..., \xi_{n-1})
\end{equation}
 for a distribution $W_{n-1}\in {\cal S}^\prime(\mathbb{R}^{d(n-1)})$, with $\xi_1=x_1-x_2$,..., $\xi_{n-1}=x_{n-1}-x_n$. Because of their spectral properties, the functions $W_{n-1}(\xi_1,..., \xi_{n-1})$ can be continued analytically to the forward tube ${\cal T}_{n-1}$. This is formed by the arrays $(\zeta_1,\zeta_2,...,\zeta_{n-1})$ of $(n-1)$ complex vectors $\zeta_j=\xi_j-i \eta_j$, where $\xi_j$ and $\eta_j$ are real vectors in $\mathbb{R}^d$ and $\eta_j\in V^+$ for $j=1,...,n-1$. The analytic continuation is done by the Laplace transform
\begin{equation}
W_{n-1}(\zeta_1,...,\zeta_{n-1}) 
= (2\pi)^{-d(n-1)}\int dp_1...dp_{n-1}\, e^{-i \sum_{j=1}^{n-1} p_j (\xi_j-i \eta_j)} \tilde{W}_{n-1}(p_1,...,p_{n-1}). 
\end{equation}
Here $\tilde{W}$ is the Fourier transform of $W$, defined as
\begin{equation}
\tilde{W}_{n-1}(p_1,...,p_{n-1})=\int d\xi_1...d\xi_{n-1}\,e^{i \sum_{j=1}^{n-1} p_j \xi_j}
W_{n-1}(\xi_1,...,\xi_{n-1})\,.
\end{equation}
The domain of analyticity can be augmented further to the extended tube ${\cal T}^{\prime (n-1)}$ by prolonging Lorentz covariance to proper complex Lorentz transformations, that is, to the group of complex matrices $\Lambda$ of unit determinant satisfying $\Lambda^T G \Lambda=G$ \cite{sw}. For a scalar field we have 
\begin{equation}
W_{n-1}(\zeta_1,..., \zeta_{n-1})=W_{n-1}(\Lambda\zeta_1,..., \Lambda\zeta_{n-1})\,.\label{lorr}
\end{equation}  
The extended tube includes in particular all the Jost points, which are the arrays $(\xi_1,...,\xi_{n-1})$ of real vectors which are all included in some wedge $\mathbb{W}^\prime$, a Lorentz transform of $\mathbb{W}$, $\mathbb{W}^\prime= \Lambda \mathbb{W}$ with $\Lambda$ any proper or unproper Lorentz transformation \cite{sw,jos}.
In particular the case $x_1\triangleleft x_2...\triangleleft x_n$ gives $\xi_j=x_j-x_{j+1}\in -\mathbb{W}$, and $(\xi_1,...,\xi_{n-1})$ is a Jost point. Consequently  ${\cal W}_n(x_1,x_2,...,x_n)=W_{n-1}(\xi_1,..., \xi_{n-1})$ is an analytic function for $x_1\triangleleft...\triangleleft x_n$.

Consider the complex Lorentz transformation $x^\prime=\Lambda_0\, x$ which leaves the coordinates $x^{2 \prime }=x^2$,..., $x^{(d-1)\prime }=x^{(d-1)}$ invariant and transforms the first two coordinates as  $x^{0 \prime }=i x^1$, $x^{1 \prime }= i x^0$. That is,
\begin{equation}
\Lambda_0=
\left(\begin{array}{cccc}
0 & i &  \ldots & 0 \\
i & 0  & \ldots & 0 \\
\vdots  & \vdots &\ddots & 0\\
0  & 0& 0& 1 
\end{array}\right)\,.
\end{equation} 
This transforms a vector $\xi \in -\mathbb{W}$, $\xi^1 <-\vert\xi^0\vert$, into a vector in ${\cal T}^1$, since
\begin{equation}
\Lambda_0 \xi= \Lambda_0 (\xi^0 , \xi^1, \xi^2,  ... , \xi^{d-1})=(i\xi^1 , i \xi^0, \xi^2, ... , \xi^{d-1}) \,.\label{termo}
\end{equation}
 This is, $\Lambda_0 \xi=\xi^\prime -i \xi^{\prime\prime}$, with $\xi^\prime=(0,0,\xi^2,...,\xi^n)$ and $\xi^{\prime\prime}=(-\xi^1 , - \xi^0, 0,  ... , 0) \in V^+$.
 
Then we can write for a Wightman 
function on  $x_1\triangleleft x_2...\triangleleft x_n$, 
\begin{eqnarray}
 {\cal W}_n(x_1,x_2,...,x_n)&=&W_{n-1}(\Lambda_0\xi_1,..., \Lambda_0\xi_{n-1}) = W_{n-1}(\xi_1^\prime-i \xi^{\prime\prime}_1,...,\xi_n^\prime-i \xi^{\prime\prime}_n)\label{parenq}\\
 &&= (2\pi)^{-d(n-1)}\int dp_1...dp_{n-1}\, e^{-i \sum_{j=1}^{n-1} p_j (\xi_j^\prime-i\xi^{\prime\prime}_j)} \tilde{W}_{n-1}(p_1,...,p_{n-1})\,.\nonumber
\end{eqnarray}
The Wightman function of the left hand side can be understood as a distribution in ${\cal S}^\prime_\triangleleft(\mathbb{R}^{dn})$, which result from a restriction of the Wightman distribution in ${\cal S}^\prime(\mathbb{R}^{dn})$ to the test functions in ${\cal S}_\triangleleft(\mathbb{R}^{dn})$.

 For a sequence of test functions $f_0$, $f_1(x_1)$,..., $f_k(x_1,...,x_k)$, with $f_j\in {\cal S}_+(\mathbb{R}^{dj})$, the left hand side of the WRP relation (\ref{wedgereflection}), can be written according to (\ref{parenq}) 
\begin{eqnarray}
&& \sum_{i,j=0}^k \int \,dx_1 ...dx_i \, dy_1... dy_j \, (f_i(\bar{x}_i,...,\bar{x}_1))^* {\cal W}_{i+j}
(x_1,...,x_i,y_1,...,y_j) f_j(y_1,...,y_j)  \nonumber \\
&&=(2\pi)^{-d(i+j-1)}\sum_{i,j=0}^k \int \,d\xi_1 ...d\xi_{i+j-1} dx_i \,(f^{-}_i(-\bar{x}_i,-\bar{\xi}_{i-1},..., -\bar{\xi}_{1}))^* f^-_j(\xi_i-x_i,\xi_{i+1},..., \xi_{i+j-1}) \nonumber \\
&&\hspace{4cm}\times \int dp_1...dp_{i+j-1}\,e^{-i \sum_{q=1}^{i+j-1} p_q (\xi_q^\prime-i\xi^{\prime\prime}_q)}\tilde{W}_{i+j-1}(p_1,...,p_{i+j-1})\,,\label{meni}
\end{eqnarray}
where we have defined $
 f^-_n(\chi_1,..., \chi_{n})=f_n(x_1,...,x_n)$, $\chi_1=-x_1$, $\chi_k=x_{k-1}-x_{k}$ for $k=2,...,n$. The points $\chi_1$,...,$\chi_n\in -\mathbb{W}$. We have that for any vector $\xi$ it is $\bar{\xi^\prime}=\xi^\prime$, and $\bar{\xi^{\prime\prime}}=-\xi^{\prime\prime}$. Using this, and interchanging the order of the coordinate and momentum integrals, we can check that (\ref{meni}) becomes
\begin{equation}
 \sum_{i,j=0}^k \int \,dp_1 ... dp_{i+j-1} \, (\hat{f}_i(p_i,...,p_1))^*  \tilde{W}_{i+j-1}
(p_1,...,p_{i+j-1}) \hat{f}_j(p_i,...,p_{i+j-1})\,,\label{doce}
\end{equation}
where
\begin{equation}
 \hat{f}_n(p_1,...,p_n)=(2 \pi)^{-d (n-1/2)}\int d\chi_1  ... d\chi_n f^-_n(\chi_1,..., \chi_n) e^{-\sum_{q=1}^n(p_q \chi_q^{\prime \prime}+i p_q \chi^\prime_q )}\,.
\label{trece}
\end{equation}
The change of the order of the integrals in (\ref{meni}) for the variables $p_l^2,...,p_l^d$ and $\xi_l^2,...,\xi_l^d$ is just the definition of the Fourier transform of a distribution. For the components $p_l^0, p_l^1$ and $\xi_l^0, \xi_l^1 $ the justification comes from the same arguments as in   the Lemma 8.4 in \cite{os}. 

Note that in eq. (\ref{doce}) the distribution $\tilde{W}_{n-1}(p_1,...,p_{n-1})$ has support on $p_l\in V_c^+$ because of the spectral condition. In this domain the functions $\hat{f}_n(p_1,...,p_n)$ of (\ref{trece}) are infinitely differentiable and of fast decrease (see for example Lemma 8.2 in \cite{os}). Thus, they can be thought as restrictions of functions $\underline{\hat{f}}(p_1,...,p_n)$ in ${\cal S}(\mathbb{R}^{d n})$ to $V_c^+$ in (\ref{doce}) (see Lemma 2.1 in \cite{os1}).   
We then write the right hand side of eq. (\ref{doce}) as
\begin{equation}
\sum_{i,j=0}^k \int \,dx_1 ...dx_i \, dy_1 ... dy_j \, (\check{f}_i(x_1,...,x_i))^* {\cal W}_{i+j}
(x_i,...,x_1,y_1,...,y_j) \check{f}_j(y_1,...,y_j)\,,\label{malv}
\end{equation}
where $\check{f}_j(x_1,...,x_j)=(2 \pi)^{-d/2}\int dp_1 ... dp_j\, \underline{\hat{f}}(p_1,...,p_j)\, e^{-i p_1 x_1 +i \sum_{q=2}^{j-1} p_q(x_{q-1}-x_{q}) }$. The quantity (\ref{malv}) is positive by the standard positivity property (\ref{wig}) for the Wightman distributions. We have then finished the proof of  (\ref{wedgereflection}).  

\subsection{Fields with spin}

It is not difficult to find the changes to (\ref{wedgereflection}) which are necessary in order to allow for fields with charge or spin. Let ${\cal W}^{(\nu, \kappa)}_n(x_1,...,x_n)=W^{(\nu, \kappa)}_{n-1}(\xi_1,...,\xi_{n-1})=\langle 0 \vert \psi^{\nu_1 \kappa_1}(x_1)...\psi^{\nu_n \kappa_n}(x_n)\vert 0 \rangle $. We use  $(\nu \kappa)$ as an abbreviation of $(\nu_1...\nu_n,\kappa_1...\kappa_n)$. The $\nu_i$ represent the  index corresponding to the finite dimensional representation of the covering group of the Lorentz group for the field $\psi^{\nu_i\kappa_i}$, labeled by $\kappa_i$. The transformation law for the field reads
\begin{equation}
U(\Lambda) \psi^{\nu \kappa}(x)U(\Lambda)^{-1}=S^\kappa(\Lambda^{-1})^{\nu}_{\nu^\prime}\psi^{\nu^\prime \kappa}(\Lambda x)\,.
\end{equation}
The adjoint field  is represented as $(\psi^{\nu \kappa}(x))^\dagger=\psi^{\nu^* \kappa^*}(x)$. It is labeled by $\kappa^*$, and the corresponding representation of the (real) Lorentz group is the complex conjugate representation to the one corresponding to $\kappa$. We also write $(\bar{\nu})= (\nu_n^*...\nu_1^*)$ and $(\bar{\kappa})=(\kappa_n^*...\kappa_1^*)$, both, taking the adjoint fields and inverting the ordering of the indices.

The covariant transformation law for the Wightman distributions now reads 
\begin{equation}
W^{(\nu, \kappa)}_{n-1}(\xi_1,..., \xi_{n-1})=\sum_{\mu} S^{(\kappa)}(\Lambda^{-1})_{(\mu)}^{(\nu)}   W^{(\mu, \kappa)}_{n-1}(\Lambda\xi_1,..., \Lambda\xi_{n-1})\,,\label{piro}
\end{equation}
where we have introduced the notation $S^{(\kappa)}(\Lambda^{-1})_{(\mu)}^{(\nu)}=S^{\kappa_1}(\Lambda^{-1})_{\mu_1}^{\nu_1}...S^{\kappa_n}(\Lambda^{-1})_{\mu_n}^{\nu_n}$.
In the extended tube, this equation holds for the complex Lorentz transformations, and in particular it extends to $\Lambda_0$, where  $S^{\kappa}(\Lambda_0^{-1})^{\mu}_{\nu}$ is the matrix corresponding to the representation of the field $\kappa$ evaluated for $\Lambda_0$ in the complex Lorentz group. 

 In order to cancel these matrix factors coming from the Lorentz transformation $\Lambda_0$ (see eq. (\ref{parenq})) we have to include extra factors to (\ref{wedgereflection}). This leads to a WRP for general fields, which writes
\begin{eqnarray}
\sum_{\stackrel{i,j}{\stackrel{(\nu_i,\kappa_i)}{(\nu_j,\kappa_j)}}} \int \,dx_1 ...dx_i \, dy_1... dy_j \, (f_i^{(\nu_i, \kappa_i)}(x_1,...,x_i))^*S^{(\bar{\kappa}_i)}(\Lambda_0)_{(\bar{\mu}_{i})}^{(\bar{\nu}_{i})}
S^{(\kappa_{j})}(\Lambda_0)_{(\mu_j)}^{(\nu_j)}&& \nonumber\\
\hspace{3.cm} {\cal W}_{i+j}^{(\bar{\mu}_i\mu_j, \bar{\kappa}_i \kappa_j)}
(\bar{x}_i,...,\bar{x}_1,y_1,...,y_j) f^{(\nu_j, \kappa_j)}_j(y_1,...,y_j)&\ge& 0\,, \label{dicete} 
\end{eqnarray}
where $f^{(\nu_l \kappa_l)}_l(x_1,...,x_l)\in {\cal S}_+(\mathbb{R}^{dl})$.

In order to find the extra matrix factors more explicitly, we can write the field representations of the one dimensional subgroup of boosts in the $x_1$ direction
\begin{equation}
\Lambda(\phi)=
\left(\begin{array}{cccc}
\cosh (\phi) & \sinh (\phi) &  \ldots & 0 \\
\sinh (\phi) & \cosh (\phi)  & \ldots & 0 \\
\vdots  & \vdots &\ddots & 0\\
0  & 0& 0& 1 
\end{array}\right)\,,
\end{equation} 
 as $S^{\kappa}(\Lambda(\phi))=e^{\phi K}$, with $\phi$ the boost parameter.  Then we have $S^{\kappa}(\Lambda_0)=e^{i\frac{\pi}{2}K}$ and $S^{\kappa^*}(\Lambda_0)=e^{i\frac{\pi}{2}K^*}$, with $K^*$ the complex conjugate of the matrix $K$. 

The $S^{(\kappa_{j})}(\Lambda_0)_{(\mu_j)}^{(\nu_j)}$ can be absorbed in the test functions in (\ref{dicete}), leaving no factors for the unbarred indices, at the expense of changing the matrix factors $S^{\kappa^*}(\Lambda_0)$ for the barred indices by their squares $(S^{\kappa^*}(\Lambda_0))^2=e^{i \pi K^*}$.
This gives, for each barred index, a factor of the wedge parity $P_{\mathbb{W}}=\Lambda_0^2=\textrm{diag}(-1,-1,1...1)$ on the vector indices, a $e^{i \frac{\pi}{2} (\alpha^{1})^*}=i(\alpha^{1})^*$ for each Dirac spinor one, and $i \alpha^{1}$ for the adjoint spinors, where $\alpha^1=\gamma^0\gamma^1$ is the Dirac matrix. In two dimensions, for a field of spin $s$, transforming as  $
U(\Lambda(\phi)) \psi(x)U(\Lambda(\phi))^{-1}=e^{-s \, \phi} \psi(\Lambda(\phi) x)$, we have a factor $e^{i \pi s^*}$ on the barred indices.  

\section{WRP, TCP and the Bisognano-Wichmann theorem}
A generalized form of the WRP inequalities can be derived in a general quantum mechanical setting using the Tomita-Takesaki modular theory \cite{tt}. Given  a cyclic and separating vector state  $\vert 0 \rangle$ in a von Neumann algebra ${\cal A}$, we can define the antilinear operator $S$ by
 \begin{equation}
 S {\cal O}\vert 0 \rangle={\cal O}^\dagger \vert 0 \rangle\,,
 \end{equation}
 for any ${\cal O}\in {\cal A}$. $S$ can be decomposed as $S=J \Delta^{\frac{1}{2}}$, with $J$ antiunitary and $\Delta$ self-adjoint and positive definite.  The crucial point of the Tomita Takesaki theory is that $J$ maps the algebra ${\cal A}$ into its commutant algebra ${\cal A}^\prime$.
 One also has $\Delta \vert 0 \rangle=\vert 0 \rangle$, $J \vert 0 \rangle=\vert 0 \rangle$, $J=J^\dagger=J^{-1}$ and $J \Delta = \Delta^{-1} J$. Then it follows, for any ${\cal O}\in {\cal A}$, and writing $\bar{{\cal O}}=J{\cal O}J$ for the "reflected" operator, $\bar{{\cal O}}\in {\cal A}^\prime$,
 \begin{equation}
 \langle 0\vert  \bar{{\cal O}} {\cal O} \vert 0 \rangle=\langle 0\vert  {\cal O} J {\cal O} \vert 0 \rangle^*=\langle 0\vert {\cal O} \Delta^{\frac{1}{2}} S {\cal O} \vert 0 \rangle^*=\langle 0\vert {\cal O} \Delta^{\frac{1}{2}} {\cal O}^\dagger \vert 0 \rangle\geq 0\,.\label{rere}
 \end{equation}
 This is a general quantum mechanical reflection positivity property. 
 
 The connection with QFT is given by the Bisognano-Wichmann theorem  \cite{aa}. This gives the modular reflection $J$ corresponding to the vacuum state and the algebra ${\cal A}_{\mathbb{W}}$ generated by the operators localized in the wedge $\mathbb{W}$. Consider the theory of a hermitian scalar field in four space-time dimensions obeying the Wightman axioms (including local commutativity), which are the hypothesis of the Bisognano-Wichmann theorem.  Then $J=U(R(e_1,\pi))\Theta$, where $U(R(e_1,\pi))$ is the unitary operator corresponding to a rotation of angle $\pi$ around the $(0,1,0,0)$ axis, and $\Theta$ is the TCP operator. The modular reflection $J$ acts geometrically on the field operators as a wedge reflection $J  \phi(x) J =\phi(\bar{x})$. 
Let $K_1$ be the boost generator in the direction of the first spatial coordinate. 
  Specifically, Bisognano and Wichmann prove that for an element ${\cal O}$ of the polynomial algebra of the field  in the 
wedge
\begin{equation}
{\cal O}=\sum_{i=0}^p \int dx_1 ... dx_i  \, f_i(x_1,...,x_i) \phi(x_1)... \phi(x_i)  \,,\label{hylo}
\end{equation}
where $f_i(x_1,...,x_i)$ is a test function with support on $\mathbb{W}$, we have
\begin{equation}
 e^{ \pi K_1 } \, {\cal O}^\dagger \vert 0\rangle=J {\cal O} \, \vert 0 \rangle\,.
\end{equation}
From this relation it follows
\begin{equation}
  \langle 0  \vert J {\cal O} J {\cal O} \, \vert 0\rangle = \langle 0\vert {\cal O} J {\cal O} \vert 0 \rangle^*=   \langle 0\vert {\cal O} e^{ \pi K_1 } \, {\cal O}^\dagger 
\vert 0\rangle \ge 0\,.\label{quete}
\end{equation}
The last inequality follows from positivity of the operator $e^{ \pi K_1 }$. This also identifies $\Delta=e^{2 \pi K_1}$. The reflected operator $\bar{{\cal O}}=J {\cal O} J$ is
\begin{equation}
 J {\cal O} J=\sum_{i=0}^p \int dx_1 ... dx_i (f_i(x_1,...,x_i))^* \phi(\bar{x}_1)... \phi(\bar{x}_i)  \,.
\end{equation}
Thus, from (\ref{quete}) we have 
\begin{equation}
\sum_{i,j=1}^k \int \,dx_1...dx_i \, dy_1... dy_j \, (f_i(x_1,...,x_i))^* {\cal W}_{i+j}
(\bar{x}_1,...,\bar{x}_i,y_1,...,y_j) f_j(y_1,...,y_j)\ge 0\,.  \label{culpa}
\end{equation}
A remark about the relation of this inequality with (\ref{wedgereflection}) is in order. 
First, in (\ref{culpa}) the test functions have support inside the wedge, but there is no restriction to the ordering of the points, nor they have to be spatially separated to each other. The WRP, eq. (\ref{wedgereflection}), follows from (\ref{culpa}) for the specific case of $f_i\in {\cal S}_+(\mathbb{R}^{dn})$, and using local commutativity in order to obtain the correct ordering for the points inside the Wightman functions (note the difference in ordering between (\ref{wedgereflection}) and (\ref{culpa})). Thus, in this sense, the relation (\ref{culpa}) is stronger than (\ref{wedgereflection}). However, (\ref{wedgereflection}) follows without the need of local commutativity (LC), or weak local commutativity, which is the condition for the validity of the TCP theorem \cite{jos}.   

It is also possible to express the WRP for any spin representation and for operators with even fermion number, in terms of a relation involving a TCP operator, if the local commutativity holds. In order to write this relation in any dimension we use a version of the TCP theorem which does not involve a reflection for all space-time coordinates, but a wedge reflection \cite{cpt}. This is equivalent to the standard TCP theorem in even dimensional space-times, because both operations are related by a rotation. However, the "wedge TCP theorem" also holds in odd dimensions, where the inversion of coordinates $x\rightarrow -x$ has determinant $(-1)$ and cannot be reached continuously from the identity by the complex Lorentz transformations. 
This wedge TCP theorem follows from the same arguments as the standard one:  The matrix $P_{\mathbb{W}}$ belongs to the complex Lorentz group, and the analyticity of the Wightman functions on the Jost points implies
\begin{equation}
{\cal W}^{(\nu, \kappa)}_{n}(x_1,...x_n)=\sum_{\mu} \left(S^{(\kappa)}(P_{\mathbb{W}})^{-1}\right)^{ (\nu)}_{(\mu)}   {\cal W}^{(\mu, \kappa)}_{n}(\bar{x}_1,..., \bar{x}_{n})\label{cristina}
\end{equation}
if $x_1\triangleleft x_2...\triangleleft x_n$. Then, if LC holds, with either commuting or anticommuting fields at space-like separated points,
 eq. (\ref{cristina}) can be interpreted as the expression of the existence of a symmetry. This is
\begin{equation}
\langle 0\vert \psi^{\nu_1 \kappa_1}(x_1)... \psi^{\nu_n \kappa_n}(x_n) \vert0 \rangle=\langle 0\vert J\psi^{\nu_1 \kappa_1}(x_1)J^{-1}...\, J\psi^{\nu_n \kappa_n}(x_n)J^{-1} \vert 0\rangle^*\,,
\end{equation}
where the antiunitary operator $J$ keeps the vacuum invariant  $J\vert0\rangle=\vert0\rangle$, and transform the fields as
\begin{equation}
J \psi^{\nu \kappa}(x) J^{-1}=i^F (S^{\kappa^*}(P_{\mathbb{W}}))^{\nu}_{\mu}\psi^{\mu^* \kappa^*}(\bar{x})\,.\label{compa}
\end{equation}
Here $F=0$ for a bosonic field and $F=1$ for a fermionic one. 
We have used $(S^\kappa(P_{\mathbb{W}})^{-1})^*=S^{\kappa^*}(P_{\mathbb{W}})$. 
Compatibility of (\ref{compa}) with anticommutation relations for fermion fields implies $J^{-1}\neq J$ on the fermion sector. This is unlike the Tomita-Takesaki theory, where $J=J^{-1}$.

When ${\cal O}$ is formed by polynomials with even number of fermion fields,
\begin{equation}
{\cal O}=\sum \int dx_1...dx_l f^{(\nu_l \kappa_l)}_l(x_1,...,x_l) \psi^{\nu_l^1 \kappa_l^1}(x_1)...\psi^{\nu_l^l \kappa_l^l}(x_l)\,,
\end{equation}
 and the components of $f_l^{(\nu_l \kappa_l)}$ belong to  ${\cal S}_+(\mathbb{R}^{dl})$, we can rewrite (\ref{dicete}), using eq. (\ref{compa}) and LC,  as $\langle 0  \vert J {\cal O} J^{-1} {\cal O} \, \vert 0\rangle \ge 0$. 

\section{Final remarks}
The WRP is a positivity property of the Wightman functions at the Jost points.  We think it might be possible to prove the Wightman axioms from the properties of analyticity, covariance and WRP for a series of functions defined exclusively at the Jost points. A proof of a reconstruction theorem in a similar fashion to the one for the euclidean axiomatic system \cite{os} is under construction. In this new "mixed" axiomatic system one would retain Lorentz covariance, but have some of the features of the euclidean system. For example, the nature of the distributions at coinciding points is not relevant, and the spectrum condition would follow from the other axioms. 

We note this scheme has resemblances to some investigations in algebraic QFT, where it was found that parting from the adequately positioned wedge regions it is possible to reconstruct the whole theory \cite{reco}. Also, as mentioned in the introduction, it is a natural system in order to study whether the Renyi entanglement entropies for the vacuum state actually define field operators \cite{cc}.  These Renyi entropies are only defined for spatially separated regions, giving place to correlators only for the Jost points.  They may provide standard Wightman fields for a class of QFT defined algebraically.

\section*{Acknowledgments}
This work was partially supported by CONICET and Universidad Nacional de Cuyo, Argentina.

\end{document}